\documentclass[aps,prb,10pt,twocolumn,superscriptaddress]{revtex4-1}
\usepackage{graphicx}
\usepackage{array}
\usepackage{amsmath}
\usepackage{makecell}
\usepackage{color}
\usepackage{pbox}
\usepackage{multirow}
\usepackage{soul}
\usepackage{orcidlink}

\newcommand{\dft}{\text{\sc dft}}

\newcommand{\refsolQuotes}{``ref\hspace{.04cm}1"}
\newcommand{\refliqQuotes}{``ref\hspace{.04cm}2"}
\newcommand{\refsolQuotesSpace}{``ref\hspace{.04cm}1"{ }}
\newcommand{\refliqQuotesSpace}{``ref\hspace{.04cm}2"{ }}

\begin{document}

\title{Accelerating ab initio melting property calculations with machine learning: Application to the high entropy alloy TaVCrW}

\author{Li-Fang Zhu*\,\orcidlink{0000-0003-2740-8172}}
\affiliation{Department for Computational Materials Design, Max-Planck-Institut f$\ddot{u}$r Eisenforschung GmbH,  Max-Planck-str.1, 40237 D$\ddot{u}$sseldorf, Germany}
\affiliation{Institute for Materials Science, University of Stuttgart, Pfaffenwaldring 55, 70569 Stuttgart, Germany}

\author{Fritz K$\rm{\ddot{o}}$rmann\,\orcidlink{0000-0003-3050-6291}}
\affiliation{Department for Computational Materials Design, Max-Planck-Institut f$\ddot{u}$r Eisenforschung GmbH,  Max-Planck-str.1, 40237 D$\ddot{u}$sseldorf, Germany}
\affiliation{Institute for Materials Science, University of Stuttgart, Pfaffenwaldring 55, 70569 Stuttgart, Germany}
\affiliation{Interdisciplinary Centre for Advanced Materials Simulation (ICAMS), Ruhr-Universität Bochum, 44801, Germany}

\author{Qing Chen\,\orcidlink{0000-0002-8392-0552}}
\affiliation{Thermo-Calc Software AB, Råsundavägen 18, 16967 Solna, Sweden}

\author{Malin Selleby\,\orcidlink{0000-0001-5031-919X}}
\affiliation{Department of Materials Science and Engineering, KTH (Royal Institute of Technology), SE-100 44 Stockholm, Sweden}

\author{J$\rm{\ddot{o}}$rg Neugebauer\,\orcidlink{0000-0002-7903-2472}}
\affiliation{Department for Computational Materials Design, Max-Planck-Institut f$\ddot{u}$r Eisenforschung GmbH,  Max-Planck-str.1, 40237 D$\ddot{u}$sseldorf, Germany}

\author{Blazej Grabowski\,\orcidlink{0000-0003-4281-5665}}
\affiliation{Institute for Materials Science, University of Stuttgart, Pfaffenwaldring 55, 70569 Stuttgart, Germany}

\date{\today}

\begin{abstract}
Melting properties are critical for designing novel materials, especially for discovering high-performance, high-melting refractory materials. Experimental measurements of these properties are extremely challenging due to their high melting temperatures. Complementary theoretical predictions are, therefore, indispensable. The conventional free energy approach using density functional theory (DFT) has been a gold standard for such purposes because of its high accuracy. However, it generally involves expensive thermodynamic integration using \textit{ab initio} molecular dynamic simulations. The high computational cost makes high-throughput calculations infeasible. Here, we propose a highly efficient DFT-based method aided by a specially designed machine learning potential. As the machine learning potential can closely reproduce the \textit{ab initio} phase space, even for multi-component alloys, the costly thermodynamic integration can be fully substituted with more efficient free energy perturbation calculations. The method achieves overall savings of computational resources by 80\% compared to current alternatives. We apply the method to the high-entropy alloy TaVCrW and calculate its melting properties, including melting temperature, entropy and enthalpy of fusion, and volume change at the melting point. Additionally, the heat capacities of solid and liquid TaVCrW are calculated. The results agree reasonably with the {\sc calphad} extrapolated values.
\end{abstract}

\maketitle

\section{Introduction\label{intro}}

Discovering novel high entropy alloys (HEAs) with exceptional performance has ushered in a new era for materials design{\cite{Senkov1,Senkov2,BG}}. The melting temperature is a crucial parameter in the search for such materials. For instance, a correlation between a high melting point and elevated temperature strength has been identified in refractory complex concentrated alloys{\cite{Senkov3}}. Besides the melting temperature, other melting properties, such as enthalpy and entropy of fusion, and volume change at the melting point, are also crucial for constructing phase diagrams and developing novel materials. However, experimental measurements on these properties, even for unary refractory materials, face severe challenges due to high melting points, often resulting in very scattered experimental data, if available at all. Additionally, the vast compositional space of HEAs makes systematic experimental screening of promising candidates impractical. 

Several computational methods for melting point predictions have been developed using empirical potentials, machine learning potentials, and density functional theory (DFT)~\cite{zhu3,MLIP2,HQJ}. Calculations on other melting properties, such as entropy and enthalpy of fusion, volume expansion from solid to liquid at the melting point, and thermodynamic properties of the liquid phase (especially the liquid heat capacity), are, however, limited. They require access to the free energy surface of both solid and liquid, including all relevant physical contributions, such as vibrational entropy, including the anharmonic contribution, and electronic entropy, including the electron-vibration coupling. These physical contributions significantly impact the thermodynamic properties of both the solid and liquid phases, thereby affecting the predicted melting properties~\cite{Moriarty,Zhang,zhu4}. 

One approach, capable of including these contributions and considered a gold standard for such calculations due to its high accuracy, is the \textit{ab initio} free energy approach within the DFT framework~\cite{Sugino,Wijs,AlfeFeNature,AlfeFePRB,AlfeAlFE,zhu1,zhu2}. In this approach, Gibbs energies of the solid and liquid phases are explicitly calculated, and the crossing point of the solid and liquid Gibbs energies determines the melting point. The solid and liquid thermodynamic properties can also be extracted from the Gibbs energies. However, achieving a high accuracy entails a significant computational cost, primarily due to the typically involved expensive \textit{ab initio} molecular dynamics (MD) simulations.\par

Methods for speeding up \textit{ab initio} solid free energy calculations have been under active development. A recent advancement in this direction is based on dedicated machine learning potentials to entirely avoid the costly \textit{ab initio} MD calculations~\cite{Duff2,BG2}. The \textit{ab initio} accuracy is then achieved by performing static DFT calculations on a few snapshots generated by the machine learning potential using the free energy perturbation theory~\cite{FEP}. This approach is called "direct upsampling" and has shown significantly improved computational efficiency compared to the previous \textit{ab initio} MD-based approaches using thermodynamic integration~\cite{BG,UP-TILD,Duff}.

For \textit{ab initio} liquid free energy calculations, due to the randomly distributed positions of the atoms, a static lattice reference is missing. Therefore, the liquid free energy calculation relies on designing a good reference to closely reproduce the \textit{ab initio} liquid phase space and perform thermodynamic integration calculations based on \textit{ab initio} MD. A recent development for speeding up the liquid free energy calculations is the TOR-TILD (\textit{two-optimized reference thermodynamic integration using Langevin dynamics}) methodology~\cite{zhu1}. It has demonstrated remarkable efficiency for calculating elemental materials and binary alloys~\cite{zhu1,zhu2,zhu4} by employing two specially designed EAM potentials as references. However, when going to multi-component alloys, its computational speed significantly slows down. The reason is that the large compositional space of multi-component alloys results in many different atomic structures. The requirements to fit an efficient reference to reach the same accuracy as for unary systems are thus more critical. In this case, EAM potentials loose their power to accurately describe the \textit{ab initio} liquid system.

Here, we propose an efficient \textit{ab initio} based approach for the liquid free energy calculation benefiting from a specially designed machine learning potential. This potential can be any machine learning model as long as it shows excellent performance regarding accuracy and efficiency, e.g., recently developed atomic cluster expansion (ACE) potentials~\cite{ACE2} and moment tensor potentials (MTPs)~\cite{MTP,MTP2}. In the present work, MTP is used. As the specially designed MTP closely overlaps with the \textit{ab initio} liquid system even for multi-component materials, it allows us to entirely skip the expensive \textit{ab initio} MD simulations and directly up-sample to the DFT level using a few snapshots from the MTP generated trajectories based on free energy perturbation.

We use the HEA TaVCrW as a prototype system to demonstrate the performance of the liquid free energy calculations. We compare them with results obtained from the conventional TOR-TILD using two EAM potentials and hybrid potentials (one EAM potential and one MTP). The free energy and thermodynamic properties of solid TaVCrW are computed using the aforementioned direct upsampling approach. Combining the thermodynamic properties of solid and liquid TaVCrW, the melting properties of TaVCrW are obtained and compared to available {\sc calphad} data, including the heat capacity and bulk modulus of the liquid phase. GGA-PBE and LDA exchange-correlation functionals are used, and their performance on melting property calculations of the TaVCrW alloy is discussed.

\begin{figure*}
\centering
\includegraphics[width=1.99\columnwidth]{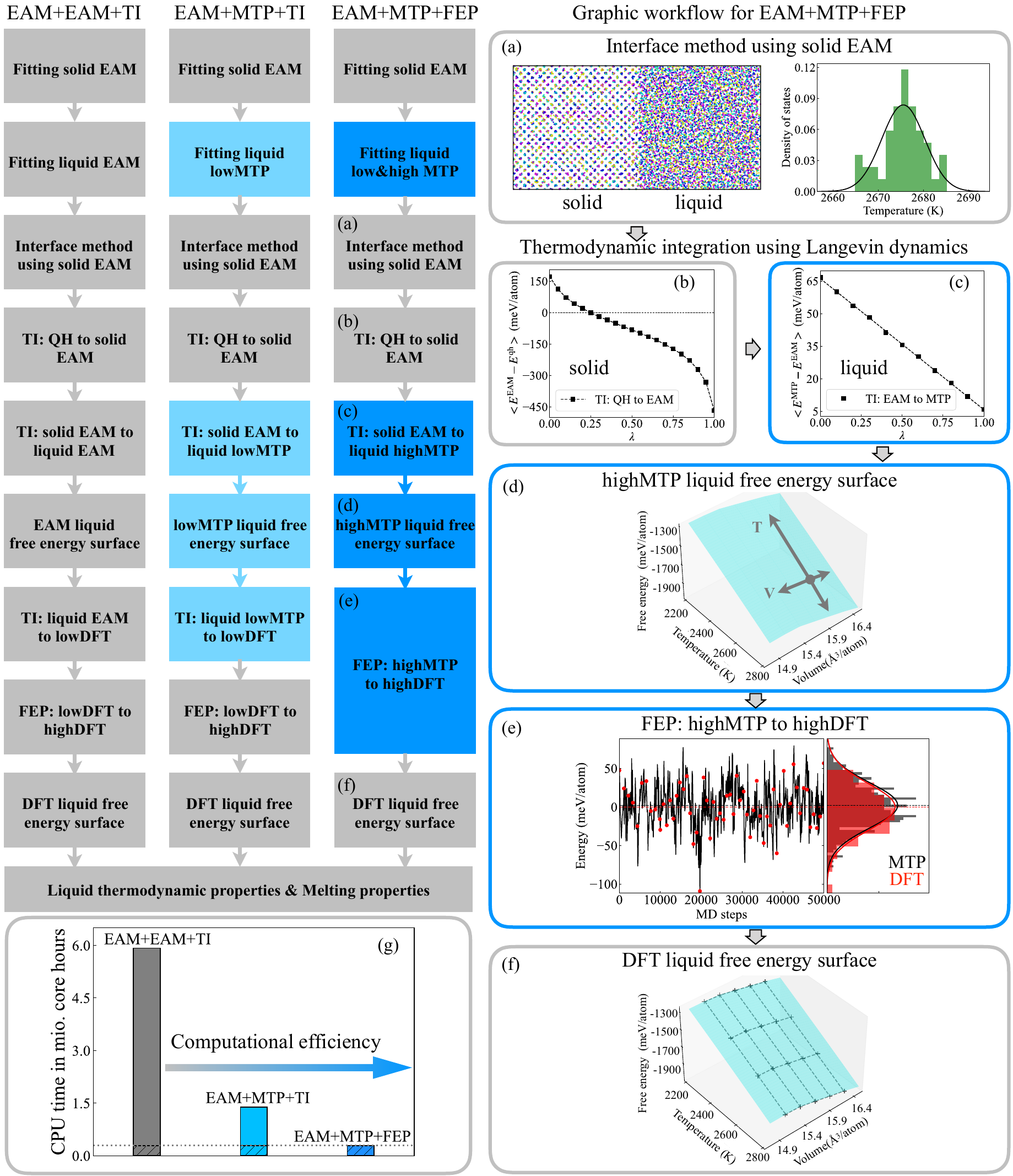}
\caption{\textbf{Schematic descriptions for liquid free energy calculations using EAM+EAM+TI, EAM+MTP+TI, and the newly proposed EAM+MTP+FEP}. Developments in EAM+MTP+TI and EAM+MTP+FEP compared to EAM+EAM+TI are marked in light and dark blue, respectively. The graphic workflow for the new EAM+MTP+FEP approach is detailed from (a) to (f). (a) shows the solid-liquid interface structure used for predicting the melting point of EAM and the Gaussian distribution of the predicted melting points. (b) and (c) depict thermodynamic integration using Langevin dynamics from effective QH to EAM and from EAM to MTP at the predicted melting point. (d) represents the liquid free energy surface of MTP calculated via thermodynamic integration from the free energy (grey dot) obtained in (c). (e) displays the MD trajectory of MTP (black line) and DFT data points (red dots) with their Gaussian distributions and mean values (dashed lines). (f) is the DFT liquid free energy surface with the DFT data points marked by black crosses. (g) compares the computational efficiency of EAM+MTP+FEP with EAM+EAM+TI and EAM+MTP+TI.\label{fig:master}}
\end{figure*}

\section{Results\label{results}}
\subsection{General overview of the methodology\label{overview}}
To access the melting properties within the free energy approach, the free energy of both solid and liquid needs to be calculated. The methodology for solid free energy calculations is mature and has been systematically investigated and documented~\cite{BG,UP-TILD,Duff,Duff2,BG2}. Here we focus on an overview of liquid free energy calculations.\par

The difficulty in calculating the liquid free energy lies in the absence of a static reference lattice. Therefore, the liquid calculation heavily depends on a reliable reference, with which the expression of the liquid free energy reads as
\begin{equation}
\begin{aligned}
F^{\rm liquid}(V,T)\,=&\,F^{\rm liquid}_{\rm ref}(V,T)+\,{\Delta}F^{\rm liquid}_{\rm ref{\to}\dft}(V,T).
\end{aligned}
\label{eq:liquid1}
\end{equation}
Here $F^{\rm liquid}_{\rm ref}(V,T)$ is the free energy of the reference and ${\Delta}F^{\rm liquid}_{\rm ref{\to}\dft}(V,T)$ the energy difference between the reference and the \textit{ab initio} liquid. ${\Delta}F^{\rm liquid}_{\rm ref{\to}\dft}(V,T)$ is the most computationally expensive part and can be obtained either by thermodynamic integration using 
\begin{equation}
\begin{aligned}
{\Delta}F^{\rm liquid}_{\rm ref{\to}\dft}\,=\int^{1}_{0}\langle E_{\rm DFT} - E_{\rm ref}\rangle_{\lambda}d\lambda,
\end{aligned}
\label{eq:TI}
\end{equation}
where $\langle \ldots \rangle_{\lambda}$ implies a thermal average on a mixed potential $E_{\lambda}$ = $\lambda$$E_{\rm DFT}$ + (1-$\lambda$)$E_{\rm ref}$, or by free energy perturbation using 
\begin{equation}
\begin{aligned}
{\Delta}F^{\rm liquid}_{\rm ref{\to}\dft}\,=-{k_{\rm B}T}{\rm ln}\langle e^{-\frac{1}{k_{\rm B}T}(E_{\rm DFT}-E_{\rm ref})}\rangle_{\lambda=0}.
\end{aligned}
\label{eq:FEP}
\end{equation}
The differences between the two techniques are that thermodynamic integration involves expensive \textit{ab initio} MD simulations, and the reference only influences the computational efficiency but does not affect the accuracy of thermodynamic integration. The accuracy of thermodynamic integration can always be reached by sufficient $\lambda$ values and \textit{ab initio} MD steps. In contrast, free energy perturbation needs only cheap reference potential calculations. Still, it is more demanding concerning the reference and works only well when the reference closely overlaps with the \textit{ab initio} system. Even though free energy perturbation could be more efficient than thermodynamic integration and could reach the same level of \textit{ab initio} accuracy as thermodynamic integration, it has not yet been widely and directly used for such calculations due to a lack of good references. Instead, thermodynamic integration has been mainly employed in the past decades.

As discussed above, the reference quality strongly influences the efficiency of both thermodynamic integration and free energy perturbation, and even the accuracy of free energy perturbation. Two fundamental factors determine the quality of the reference: firstly, it should closely resemble the \textit{ab initio} liquid, and secondly, its free energy must be either known or calculable. However, meeting both of these criteria simultaneously proves challenging.\par

To solve this challenge, Alf{\`e} $et$ $al$. proposed a solution in the early stages, suggesting the combination of a Lennard-Jones potential with an Invert-Power (IP) potential for computing the liquid free energy of Fe and Al~\cite{AlfeFeNature,AlfeFePRB,AlfeAlFE}. The IP potential, being closer to the \textit{ab initio} liquid, serves as a reference for performing thermodynamic integration to the \textit{ab initio} liquid. Calculating the free energy of the IP potential involves employing thermodynamic integration from the Lennard-Jones potential, as the liquid free energy of the Lennard-Jones potential has already been tabulated. A recent advancement that significantly enhances computational efficiency and accuracy of liquid free energy calculation is the aforementioned TOR-TILD methodology~\cite{zhu1} using two specially designed EAM potentials as references. Here, the first EAM is solely for obtaining the free energy of the reference, and the second EAM is for expediting thermodynamic integration calculations. Under these conditions, the liquid free energy can be reformulated as
\begin{equation}
\begin{aligned}
F^{\rm liquid}(V,T)\,=&\,F^{\rm liquid}_{\rm ref1}(V,T)+{\Delta}F^{\rm liquid}_{\rm ref1{\to} ref2}(V,T)\\&+\,{\Delta}F^{\rm liquid}_{\rm ref2{\to}\dft}(V,T).
\end{aligned}
\label{eq:liquid2}
\end{equation}
Here, $F^{\rm liquid}_{\rm ref1}$ is the free energy of the first reference fitted to the \textit{ab initio} solid energies (labeled \refsolQuotes), ${\Delta}F^{\rm liquid}_{\rm ref1{\to}ref2}$ the free energy difference between \refsolQuotesSpace and the second reference fitted to the \textit{ab initio} liquid energies (labeled \refliqQuotes), and ${\Delta}F^{\rm liquid}_{\rm ref2{\to}\dft}$ the free energy difference between \refliqQuotesSpace and the \textit{ab initio} liquid, similar as ${\Delta}F^{\rm liquid}_{\rm ref{\to}\dft}$ in Eq.~(\ref{eq:liquid1}). Note that the vibrational entropy and electron-vibration coupling are fully included during the calculations of ${\Delta}F^{\rm liquid}_{\rm ref2{\to}\dft}$. \par

Within the TOR-TILD methodology, determining the first two terms in Eq.~(\ref{eq:liquid2}) involves only EAM calculations, making it extremely efficient~\cite{zhu4,zhu1,zhu2}. However, addressing the challenges of EAMs related to multi-component materials, as mentioned in Sec.~\ref{intro}, requires using machine learning potentials instead of EAM. The primary challenge is that MD calculations using machine learning potentials, here MTP, are generally 30-50 times slower than those using EAM. Simply substituting EAM with MTP for the large set of MD calculations required for the first two terms in Eq.~(\ref{eq:liquid2}) demands significantly more computing effort. Therefore, an optimal solution is to employ a hybrid potential approach using EAM as \refsolQuotesSpace and MTP as \refliqQuotesSpace. Here, due to the high quality of MTP, the DFT accuracy can be reached by performing efficient free energy perturbation calculations instead of costly thermodynamic integration calculations.\par

To address the efficiency of the proposed EAM+MTP+FEP approach, the schematic descriptions of the conventional TOR-TILD approach using two EAM potentials (EAM+EAM+TI), and using one EAM potential and one MTP (EAM+MTP+TI) are also given in Fig.~\ref{fig:master}. Compared to EAM+EAM+TI, the new developments are marked as light blue in EAM+MTP+TI and as dark blue in EAM+MTP+FEP. The graphic workflow for EAM+MTP+FEP is specifically demonstrated by Fig.~\ref{fig:master} (a)-(f). Before performing this workflow, an EAM potential is independently fitted as \refsolQuotesSpace based on solid DFT energies and an MTP as \refliqQuotesSpace based on liquid DFT energies. The fitting details are provided in Sec.~\ref{PotentialFitting}.\par

Within the workflow, the melting point of \refsolQuotes, $T^{\rm m}_{\rm ref1}$, is firstly computed through the interface method~\cite{MorrisInterfaceMethod} based on MD simulations. Different random seeds result in slightly different melting points, so the Gaussian mean value of a few tens of predicted melting points is used for the following step, as shown by Fig.~\ref{fig:master} (a). The solid free energy at $T^{\rm m}_{\rm ref1}$ and corresponding equilibrium volume, $F^{\rm solid}_{\rm ref1}$($V^{\rm m,solid}_{\rm ref1}$,$T^{\rm m}_{\rm ref1}$), is then calculated through thermodynamic integration at different $\lambda$ values from an effective quasiharmonic (QH) reference (see Fig.~\ref{fig:master} (b)). For details on calculating the free energy of the effective QH, we refer to Ref.~[\onlinecite{Duff2}]. Since at the melting point the Gibbs energies of solid and liquid are equal at constant pressure (here $P$\,=\,0\,GPa), we can easily relate the Helmholtz free energies of the solid and liquid at the melting point via $G(P,V)$\,=\,$F(P,V)$\,+\,$PV$, i.e., $F^{\rm solid}_{\rm ref1}$($V^{\rm m,solid}_{\rm ref1}$,$T^{\rm m}_{\rm ref1}$)\,=\,$F^{\rm liquid}_{\rm ref1}$($V^{\rm m,liquid}_{\rm ref1}$,$T^{\rm m}_{\rm ref1}$). Note that $V^{\rm m,solid}_{\rm ref1}$ is not equal to $V^{\rm m,liquid}_{\rm ref1}$ due to the volume expansion from solid to liquid at the melting point. $F^{\rm liquid}_{\rm ref1}$($V^{\rm m,liquid}_{\rm ref1}$,$T^{\rm m}_{\rm ref1}$) then acts as a reference to access $F^{\rm liquid}_{\rm ref2}$($V^{\rm m,liquid}_{\rm ref1}$,$T^{\rm m}_{\rm ref1}$) through thermodynamic integration again from \refsolQuotesSpace to \refliqQuotes (see Fig.~\ref{fig:master} (c)). The output of Fig.~\ref{fig:master} (c) is $F^{\rm liquid}_{\rm ref2}$($V^{\rm m,liquid}_{\rm ref1}$,$T^{\rm m}_{\rm ref1}$) and is shown as a grey dot in Fig.~\ref{fig:master} (d). Here, MD simulations using \refliqQuotesSpace at a dense set of volumes and temperatures are performed. By integrating the pressure along the volume dimension and the internal energy along the temperature dimension, the full free energy surface of \refliqQuotesSpace is obtained as shown by the cyan surface in Fig.~\ref{fig:master} (d). To access the \textit{ab initio} accuracy, a set of uncorrelated snapshots from the MTP trajectories (black line in Fig.~\ref{fig:master} (e)) are computed with static DFT calculations (the red dots in Fig.~\ref{fig:master} (e)). Note that, in the conventional TOR-TILD, the trajectories are from expensive \textit{ab initio} MD calculations. The small energy difference of 2 meV/atom between MTP and DFT, shown by the black and red dashed lines in Fig.~\ref{fig:master} (e), is added on top of the free energy surface of MTP. Afterward, the full liquid free energy surface from \textit{at initio} is obtained, as shown in Fig.~\ref{fig:master} (f) where the black crosses are the specific data points calculated from DFT. The thermodynamic properties of liquid TaVCrW are extracted based on the liquid free energy surface. Combining the solid thermodynamic properties, the melting properties of TaVCrW can be accessed; see Fig.~\ref{fig:PBE} and Fig.~\ref{fig:LDA}. Note that the GGA-PBE solid data are from Ref.~[\onlinecite{Duff2}], the GGA-PBE liquid data, and the LDA solid and liquid data are calculated in the present work. More details about the workflow can be found in Sec.~\ref{method_detail}.\par

\begin{figure*}
\centering
\includegraphics[width=1.99\columnwidth]{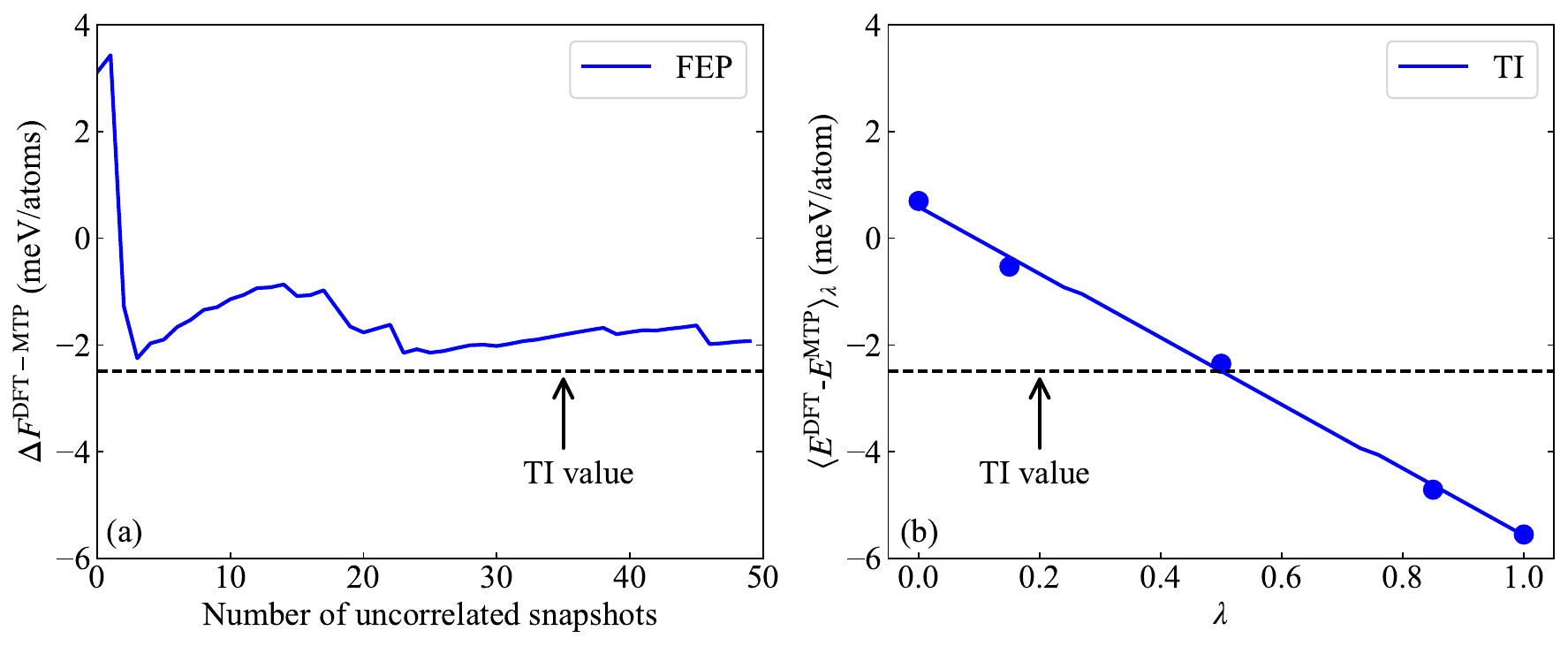}
\caption{\textbf{Free energy difference between MTP and DFT using GGA-PBE for liquid TaVCrW at $T$ = 2400\,K and $V$ = 16.24 \r{A}$^{3}$/atom.} (a) Calculated from free energy perturbation (FEP). (b) Calculated from thermodynamic integration (TI). The TI value is the free energy difference computed as the area under the blue line in (b), which is a linear fit of the five blue dots. The resulting energy difference between the free energy perturbation and the thermodynamic integration approach is only 0.5 meV/atom.
\label{fig:compareFEPTI}}
\end{figure*}

\subsection{Efficiency and accuracy\label{efficiencyANDaccuracy}}

\begin{table}
\begin{ruledtabular}
\caption{\label{tab:CPU}CPU time in core hours needed by three combinations, I. EAM+EAM+TI, II. EAM+MTP+TI and III. EAM+MTP+FEP, for computing the \textit{ab initio} free energy surface of liquid TaCrVW with 5$\times$6 ($V$, $T$) grid using LDA. The time for fitting potentials are also listed and included in the total time. Step 1 to Step 5 are defined in Sec.~\ref{steps}.}
\begin{tabular}{ccccc}
&&I.&II.&III.\\[.1cm]\hline\\[-.25cm]
&$t$(solid EAM)&15,360&15,360&15,360\\
potential&$t$(liquid EAM)&15,360&-&-\\
fitting&$t$(liquid lowMTP)&-&15,360&15,360\\
&$t$(liquid highMTP)&-&-&36,000\\[.1cm]\hline\\[-.25cm]
&$t$(Step 1)&630&630&630\\
&$t$(Step 2)&25&110&110\\
free energy&$t$(Step 3)&52&2,319&2,319\\
calculation&$t$(Step 4)&5,526,720&990,960&23\\
&$t$(Step 5)&360,000&360,000&216,000\\[.1cm]\hline\\[-.25cm]
&$t$(sum)&5,918,147&1,384,739&285,802\\[.1cm]
\end{tabular}
\end{ruledtabular}
\vspace{.5cm}
\end{table}

The key of the proposed approach is the highly optimized MTP that provides a strong overlap to the phase space sampling of the \textit{ab initio} liquid (see Fig.~\ref{fig:master} (e)). This overlap enables more efficient free energy perturbation calculations and avoids the computationally expensive thermodynamic integration calculations based on \textit{ab initio} MD. To achieve DFT accuracy, the energy difference between the MTP and the \textit{ab initio} liquid phase can be computed using Eq.~(\ref{eq:FEP}), leveraging a few uncorrelated snapshots from the MD trajectories of the MTP. The number of required snapshots depends on the root-mean-square error (RMSE) of the MTP, following the relation (2RMSE/$c$)$^{2}$, where $\pm$$c$ represents the target accuracy. This relationship has been tested and discussed in detail in Ref.~[\onlinecite{BG2}]. According to the RMSE of the MTP (here 2.6 meV/atom for GGA-PBE calculations), to attain an accuracy of 1 meV/atom, only approximately 28 uncorrelated snapshots are needed, as depicted in Fig.~\ref{fig:compareFEPTI} (a).

We compare the computational effort of the proposed EAM+MTP+FEP approach using free energy perturbation calculations with two other approaches within the thermodynamic integration framework using the conventional TOR-TILD: EAM+EAM+TI and EAM+MTP+TI. To have consistent comparisons, \refsolQuotesSpace in the three approaches are fitted to solid DFT energies with low-converged DFT parameters; \refliqQuotesSpace in EAM+EAM+TI and EAM+MTP+TI are fitted to liquid DFT energies with low-converged DFT parameters; and \refliqQuotesSpace in EAM+MTP+FEP is fitted to liquid DFT energies with high-converged DFT parameters. The required CPU core hours for the three approaches in fitting potentials and computing a liquid free energy surface with a 5$\times$6 ($V$,$T$) grid are listed in Table~\ref{tab:CPU}, following the steps described in Sec.~\ref{steps}. Compared to EAM+EAM+TI, the proposed EAM+MTP+FEP approach is about 20 times faster in total CPU time. This substantial benefit stems from a significant reduction in CPU core hours in Step 4 on computing the black trajectories in Fig.~\ref{fig:master} (e), from 5,526,720 to only 23 CPU core hours, even though the computing efficiency experiences a slight slowdown in Step 2 (Fig.~\ref{fig:master} (c)) and Step 3 (Fig.~\ref{fig:master} (d)) where MTP calculations are involved. This acceleration is attributed to substituting EAM by MTP and TI by FEP. When comparing EAM+MTP+FEP with EAM+MTP+TI, the required computational resources are saved by about 80$\%$ in calculating a 5$\times$6 ($V$,$T$) free energy surface. Note that the saving factor increases with the number of sampling points ($V$,$T$). Especially for accurately determining liquid heat capacity, a denser free energy surface is generally required, as the heat capacity is related to the second derivative of the Gibbs energy with respect to temperature. The computational efficiency of these three approaches is also illustrated in Fig.~\ref{fig:master} (g).

The proposed methodology can achieve the same level of \textit{ab initio} accuracy as the conventional TOR-TILD. The energy differences between the MTP and DFT computed by free energy perturbation and thermodynamic integration are relatively small (0.5 meV/atom), as shown in Fig.~\ref{fig:compareFEPTI} (a) and (b). They result
from the chosen convergence accuracy of 1 meV/atom for the thermodynamic sampling in thermodynamic integration and free energy perturbation. These small energy differences result in a minor melting point shift of 15\,K.

\subsection{Electronic contribution and its coupling to vibration\label{electronic}}

\begin{figure}
\centering
\includegraphics[width=0.99\columnwidth]{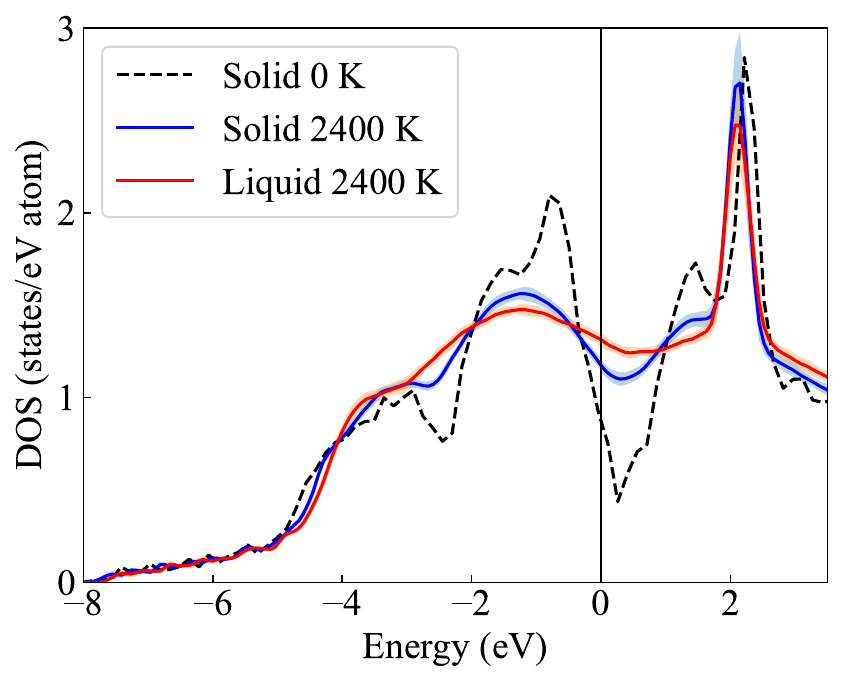}
\caption{\textbf{The electronic density of states (DOS) of TaVCrW}. The blue/red line is for solid/liquid at 2400\,K and respective equilibrium volumes including electron-vibration coupling. They are calculated from a statistically converged set of uncorrelated snapshots. The blue/red shaded areas show the standard deviation from the different snapshots. The black dashed line is for the ideal static lattice at 0\,K with the consistent electronic temperature. All DOSs shown here are computed using GGA-PBE.\label{fig:DOS}}
\end{figure}

To study the effect of electronic excitations and their coupling to vibrations on the melting properties of TaVCrW, the DOS of the solid and liquid at 2400\,K using their respective equilibrium volumes are calculated. The GGA-PBE results are illustrated in Fig.~\ref{fig:DOS}. The DOS of an ideal static lattice at 0\,K, with a consistent electronic temperature of 2400\,K, is also provided as a reference (depicted by the black dashed line in Fig.~\ref{fig:DOS}). This representation highlights how thermal vibrations smooth out the DOS of both solid and liquid phases at elevated temperatures.

The electronic contribution to the free energy of solid and liquid differs. This disparity results from a gap of 0.16 states/eV$\cdot$atom in the DOS of solid and liquid phases at the Fermi level. Correspondingly, this gap translates to an electronic Gibbs energy difference of 4.2 meV/atom between the solid and liquid, leading to a shift in the melting point by 48\,K compared to calculations that do not account for the electronic contribution. For the LDA calculations, the electronic Gibbs energy difference between solid and liquid is about 5.8 meV/atom resulting in a melting point shift by 57\,K.

\subsection{Melting properties\label{melting}}

\begin{figure*}
\centering
\includegraphics[width=2\columnwidth]{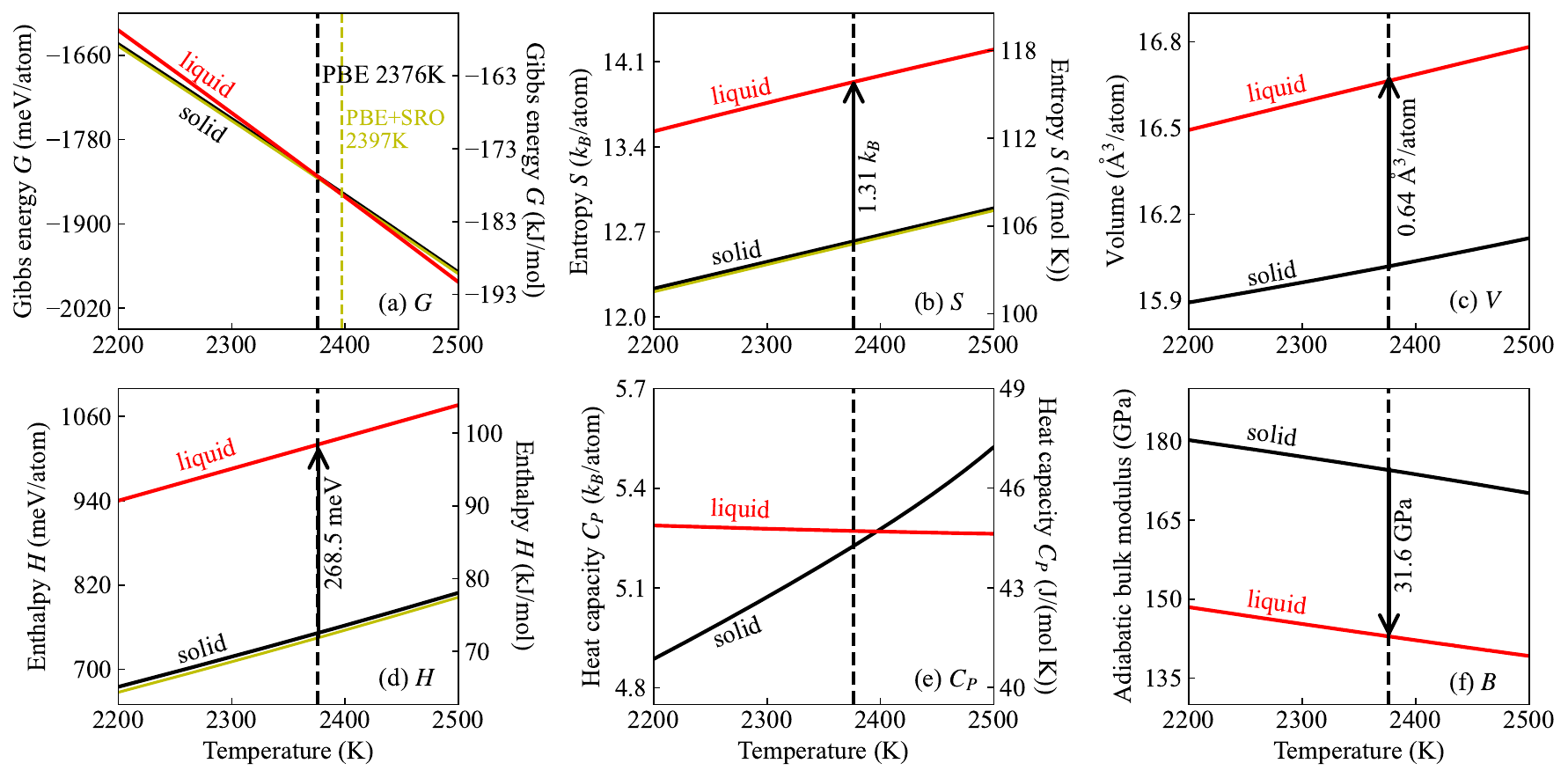}
\caption{\textbf{Temperature dependence of thermodynamic properties of solid and liquid from GGA-PBE calculations}. (a) Gibbs energy ($G$), (b) entropy ($S$), (c) volume ($V$), (d) enthalpy ($H$), (e) heat capacity ($C_P$), and (f) bulk modulus ($B$). The yellow lines in (a), (b), and (d) represent the solid Gibbs energy, entropy, and enthalpy, including the SRO contribution, which slightly stabilizes the solid phase and results in a higher melting point 2397\,K.\label{fig:PBE}}
\end{figure*}

\label{LDA}
\begin{figure*}
\centering
\includegraphics[width=2\columnwidth]{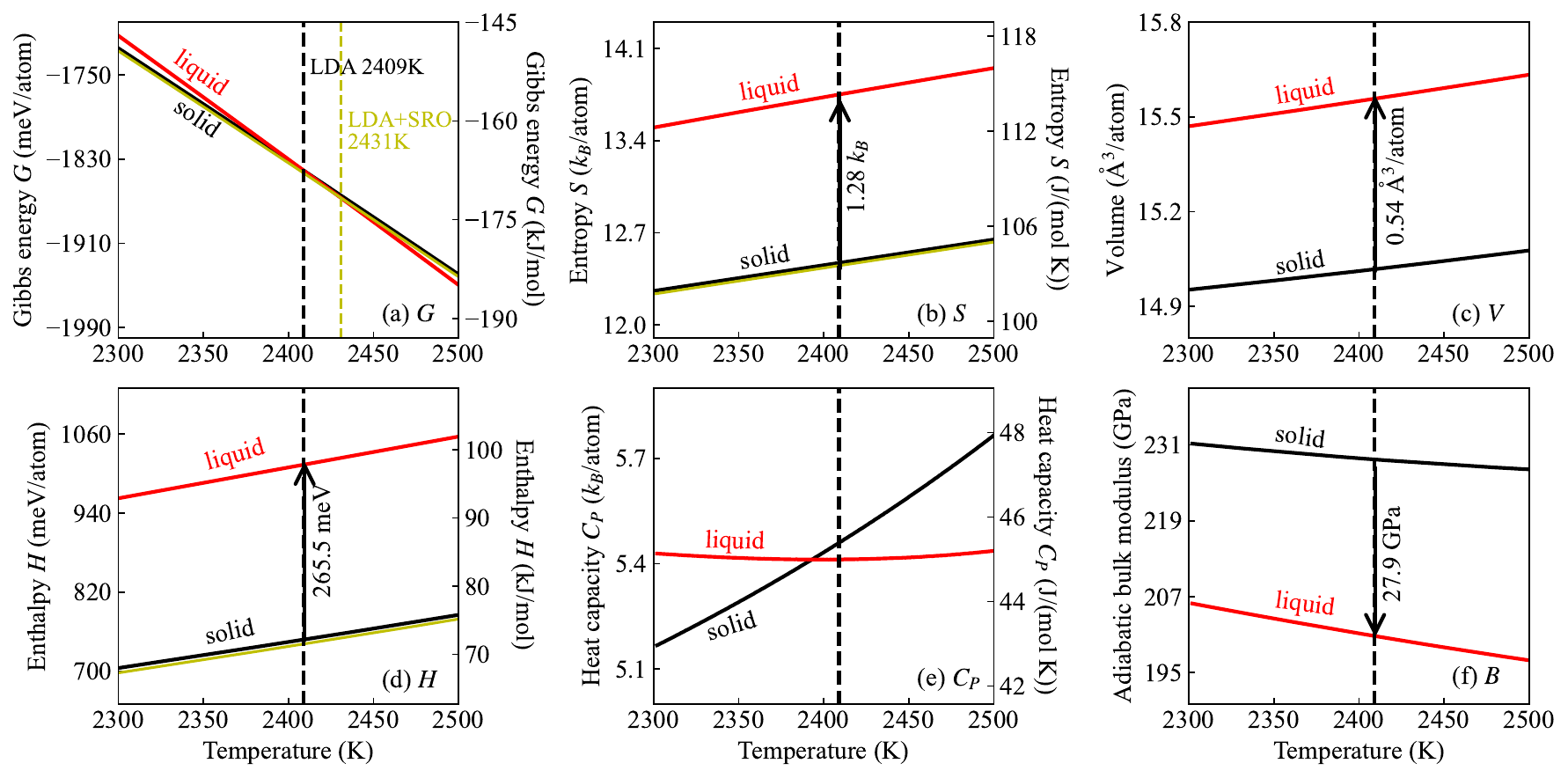}
\caption{\textbf{Temperature dependence of thermodynamic properties of solid and liquid from LDA calculations.} (a) Gibbs energy ($G$), (b) entropy ($S$), (c) volume ($V$), (d) enthalpy ($H$), (e) heat capacity ($C_{\rm P}$), and (f) bulk modulus ($B$). The yellow lines in (a), (b), and (d) represent the solid Gibbs energy, entropy, and enthalpy including the SRO contribution, which slightly stabilizes the solid phase and results in a higher melting point 2431\,K. \label{fig:LDA}}
\end{figure*}

\begin{table}
\begin{ruledtabular}
\caption{\label{tab:DetaQuantities}Melting properties of TaVCrW calculated with GGA-PBE and LDA compared to those extracted from the {\sc calphad} method.}
\begin{tabular}{ccccc}
&GGA-PBE&LDA&{\sc calphad}\\[.1cm]\hline\\[-.25cm]
$T^{\rm m}$ (K)&2376&2409&2569\\\hline\\[-.25cm]
$\Delta H^{\rm m}$ (kJ/mol)&25.9&25.6&23.8\\
$\Delta H^{\rm m}$ (meV/atom)&268.5&265.5&246.3\\[.1cm]\hline\\[-.25cm]
$\Delta S^{\rm m}$ (J/(mol K))&10.9&10.63&9.25\\
$\Delta S^{\rm m}$ ($k_{\rm B}$/atom)&1.31&1.28&1.11\\[.1cm]\hline\\[-.25cm]
$C_{P,\rm solid}^{\rm m}$ (J/(mol K))&43.45&45.39&42.58\\
$C_{P,\rm liquid}^{\rm m}$ (J/(mol K))&43.84&44.99&43.55\\[.1cm]\hline\\[-.25cm]
$\Delta V^{\rm m}$ (\AA$^{3}$/atom)&0.64&0.54&0.51\\
$V^{\rm m}_{\rm solid}$ (\AA$^{3}$/atom)&16.02&15.02&16.37\\
$V^{\rm m}_{\rm liquid}$ (\AA$^{3}$/atom)&16.66&15.56&16.88\\
\end{tabular}
\end{ruledtabular}
\vspace{.5cm}
\end{table}

Unlike unary materials, alloys typically exhibit a distinct solidus and liquidus temperature, resulting in a melting interval with upper (liquidus) and lower (solidus) limits, where both solid and liquid phases coexist. Here, the melting point we investigate is the crossing point of the solid and liquid Gibbs energy, called $T_{0}$ within the {\sc calphad} community.

As there are no experimental melting properties for TaVCrW available for a comparison with our DFT results, we extrapolate the melting properties of TaVCrW using the {\sc calphad} method with Thermo-Calc 2023a{~\cite{Thermo-Calc}}. The high entropy alloy TCHEA4 database{~\cite{TCHEA4}} is utilized, which includes only binary descriptions for all subsystems in the Ta-V-Cr-W quaternary system. This is understandable, given the lack of adequate experimental ternary phase diagram information. While no general rules exist for assessing the uncertainty of a {\sc calphad} calculation based on binary system extrapolation, the application of the TCHEA database and its predecessor, TCNI, has been successful in designing refractory high entropy alloys{~\cite{2015Zhang1,2015Zhang2,2016Gao,2018Coury,2021Wang, 2022Rao}}.

Figure~\ref{fig:PBE} and Fig.~\ref{fig:LDA} depict the temperature dependence of Gibbs energy ($G$), entropy ($S$), volume ($V$), enthalpy ($H$), heat capacity ($C_{P}$), and bulk modulus ($B$) for solid (black line) and liquid (red line) TaVCrW from GGA-PBE and LDA. The Gibbs energies are referenced to the internal energy of solid TaVCrW at $T=0$\,K. Table~\ref{tab:DetaQuantities} compiles the calculated properties from GGA-PBE and LDA and those extracted from the {\sc calphad} method. Our DFT predicted melting point from GGA-PBE and LDA are respectively 2376\,K and 2409\,K. Both values are lower than the {\sc calphad} value of 2569\,K, but still within the melting interval between 2335\,K (solidus) and 2805\,K (liquidus) extracted from the {\sc calphad} method. Our predicted enthalpy of fusion, the entropy of fusion, the heat capacity of both solid and liquid, and the volume change at the corresponding melting point are slightly larger than the {\sc calphad} values, whereas the predicted volumes for solid and liquid are smaller than the {\sc calphad} values. Note that the temperature dependence of the bulk modulus for both solid and liquid is also easily accessed with our approach, as shown in Fig.~\ref{fig:PBE} (f) and Fig.~\ref{fig:LDA} (f), where the solid bulk modulus is higher than the liquid one at the corresponding melting point by 31.6 GPa from GGA-PBE and 27.9 GPa from LDA. However, the existing database within Thermo-Calc does not contain data for bulk modulus for comparison.

Note that even though the absolute values of the thermodynamic properties for both the solid and liquid are different between GGA-PBE and LDA, the temperature dependence is similar, as shown in Fig.~\ref{fig:PBE} and Fig.~\ref{fig:LDA}, which is consistent with our previous finding for tungsten~\cite{zhu4}.

\subsection{Configurational entropy and impact of short-range order}

We assume an ideal mixing of elements for the solid and liquid free energy calculations. To quantify the potential degree of short-range order and its possible impact we performed simulations employing on-lattice machine learning potentials, namely the low-rank potentials, see Sec.~\ref{method_detail} for the technical details. We focus on the high-temperature regime of 2200 - 2500\,K as in Fig.~\ref{fig:PBE}, relevant for melting properties. To allow for efficient computation, we have chosen a temperature of 2260\,K, which is about 100\,K below the GGA-PBE computed melting temperature, to fix the volume and electronic free energy contributions. 

We computed the configurational part of the internal energy, $\Delta U^{\rm solid}_{\rm conf}$, entropy, $\Delta S^{\rm solid}_{\rm conf}$, and free energy, $\Delta F^{\rm solid}_{\rm conf}$ referenced to the ideal random alloy. These quantities allow for a direct evaluation of the impact of SRO as compared to the ideal random alloy assumption. The internal energy is directly accessible via the Monte Carlo simulations. The entropy contribution is computed by integrating the heat capacity from high temperatures downwards~\cite{PhysRevMaterials.5.053803}.

Due to SRO, the configurational entropy is slightly smaller than the ideal configurational entropy of $\ln(4)$. It, therefore, results via $-TS$ in a positive (destabilizing) entropy-driven energy contribution of around 5-6\,meV/atom. However, SRO also decreases the internal energy by about 8-9\,meV/atom (Fig.~\ref{fig:PBE} (d)). Hence the resulting free energy of the solid is overall stabilized by about 2.4\,meV/atom, Fig.~\ref{fig:PBE} (a), resulting in a slight shift of the melting temperature as discussed further below. 

\section{Discussion\label{discuss}}

A hierarchical approach for efficient calculations of the liquid free energy of high entropy alloys with \textit{ab initio} accuracy has been introduced. The approach integrates EAM, MTP, and DFT to achieve progressively increasing accuracy while maintaining maximally optimized computational speed. This combination improves the overall computational efficiency by saving 80$\%$ of the computational resources as compared to similarly accurate methods while retaining accuracy in the sub-meV range. The primary innovation lies in leveraging the advantages of MTP and replacing the expensive thermodynamic integration with free energy perturbation calculations. The substantial reduction in computing effort makes the approach attractive for high-throughput calculations of melting properties of multi-component alloys with \textit{ab initio} accuracy.

The approach has been applied to the bcc refractory high entropy alloy TaVCrW, for which currently no experimental data of melting properties exist. The results obtained with the new method closely align with those from the conventional and highly accurate TOR-TILD methodology (deviations $<$0.5\,meV/atom). To put the predicted properties into perspective, we have compared the data with those computed with the {\sc calphad} method using Thermo-Calc 2023a. Both our DFT predictions are below the {\sc calphad} extracted value (2569\,K) by 193\,K for GGA-PBE and 160\,K for LDA. A few uncertainties which may contribute to this discrepancy are listed in the following.\par 

First, for the {\sc calphad} extracted melting point, the input data in the high entropy alloy database for parametrization includes only sub-binary alloy information for TaVCrW. Although direct evidence for the uncertainty of the {\sc calphad} predicted melting point of the TaVCrW alloy is lacking, we can reasonably speculate that the error is within $\pm$100\,K, based on comparing {\sc calphad} predicted values with those from experimental and theoretical data for similar refractory alloy systems{~\cite{ZrNbTiVHf,2022Rao,2022Mishra}}.\par

Second, for the DFT calculations, the estimated error in the computed Gibbs energy arising from statistics and fitting is about 2.5 meV/atom for both the solid and liquid. For the free energy approach, an error of 1 meV/atom in either solid or liquid phase would introduce a $\sim$10\,K shift in the melting point prediction. We, therefore, estimate the maximum numerical error in the DFT predicted melting points as $\pm$50\,K.\par

Third, high-temperature magnetic fluctuations have not been included. Cr and V, for example, are treated as non-magnetic elements at the investigated temperature range. We have performed additional spin-polarized test calculations using GGA-PBE based on high-temperature MD snapshots, revealing a negligible impact. However, it was reported that standard DFT employing GGA-PBE or LDA fails to account for strong magnetic fluctuations in bcc Cr~\cite{FK}. More elaborate treatment of magnetism may resolve this issue, which is, however, beyond the scope of the present work. 

Fourth, the SRO contribution to the solid Gibbs energy stabilizes the solid phase by 2-3 meV/atom at $T_{\rm m}$. This contribution is relatively small because the melting temperature is nearly double the chemical order-disorder temperature~\cite{Damian}. The effect on the calculated melting temperature is approximately 25\,K. While SRO contributions may have a more significant impact in alloys with strong ordering and greater SRO near the melting temperature, SRO does not significantly affect the computation of the melting temperature in the current alloy. \par

Overall, considering the discussed uncertainty of the {\sc calphad} extrapolated value and the accumulated possible uncertainties from the DFT predictions, we achieve a reasonable agreement for the melting temperature value of bcc TaVCrW. 

Apart from the quantitative comparison, a crucial merit of the approach is that it provides physical insights into how different thermal contributions affect the melting properties of materials, in particular, the electronic excitations and their coupling to lattice vibrations. In our previous work~\cite{zhu4}, we observed a significant impact of electronic excitations in tungsten resulting in a large melting point shift (178\,K for GGA-PBE and 158\,K for LDA). In the present work, given that tungsten constitutes 1/4 of the investigated TaVCrW alloy, the induced change in melting point is smaller but still significant (48\,K for GGA-PBE and 57\,K for LDA) and again highlights the necessity to include the electronic contribution into melting point considerations.

\begin{figure}
\centering
\includegraphics[width=0.99\columnwidth]{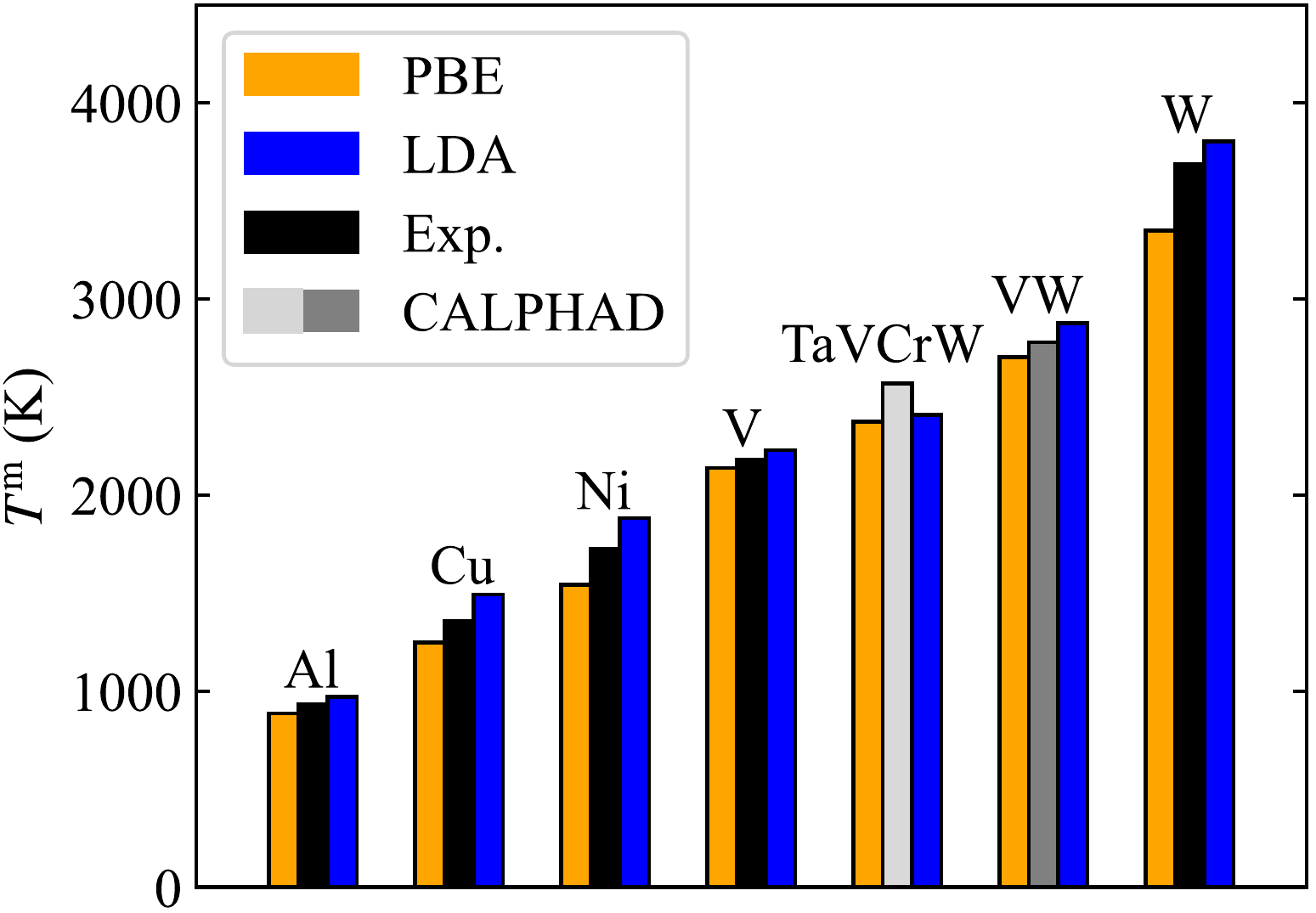}
\caption{\textbf{Performance of the standard exchange-correlation functional GGA-PBE and LDA on melting point predictions}. DFT data for Cu from Ref.~[\onlinecite{zhu1}], for Al and Ni from Ref.~[\onlinecite{zhu2}], for V, VW, and W from Ref.~[\onlinecite{zhu4}]. The black bars indicate the experimental melting points. The grey and light grey bars are the melting points extrapolated from the {\sc calphad} method. Note that explicit experimental melting data is available for binary VW (grey bar). In contrast, for TaVCrW the extrapolated melting point from the {\sc calphad} method (light grey bar) is an approximate theoretical prediction based on available binary alloys in the database. \label{fig:performance}}
\end{figure}

The efficiency of the approach allows us also to evaluate the performance of different exchange-correlation functionals on melting property predictions within the DFT framework, which would be otherwise a computationally much more challenging task. Our previous works ~\cite{zhu1,zhu2,zhu4} have shown that GGA-PBE and LDA provide an \textit{ab initio} confidence interval for the experimental melting points, i.e., GGA-PBE predicts a lower melting boundary and LDA a higher melting boundary, as shown in Fig.~\ref{fig:performance}. The reason is attributed to the underbinding / overbinding property of GGA-PBE / LDA. For TaVCrW, using the {\sc calphad} value as a proxy for experiment, this empirical rule seems to not hold anymore. However, this does not necessarily invalidate this rule but could be related to the aforementioned uncertainties in the {\sc calphad} value. To further explore the DFT empirical rules of the standard functional in this aspect, we considered their performance for the equilibrium lattice constants at $T$ = 0\,K. Already for the pure elements, the underbinding / overbinding tendency of GGA-PBE / LDA, which generally predicts larger / smaller equilibrium lattice constants compared to experimental values for most unaries,  does not hold for unary V~\cite{zhu4,Blaha,Cr1} and Cr~\cite{Cr1,Cr2} anymore, where both GGA-PBE and LDA predict smaller equilibrium lattice constants at $T$ = 0\,K. This discrepancy may propagate into the here-considered TaVCrW multi-component alloy. It would be therefore instructive to first elaborate on different methods (e.g., dynamical mean-field theory), capable of resolving the $T$ = 0\,K discrepancies for the unaries~\cite{electronicV,paramagnetismV,zhu4}, before employing these computationally more expensive approaches for melting point calculations. We also note that other DFT treatments can be integrated into the proposed free energy perturbation approach since it requires only the computation of energies but not forces. 

In summary, the proposed method opens an avenue to computationally efficient (resources savings of 80$\%$), yet highly accurate (sub-meV range) determination of melting properties of multi-component alloys. Given the high melting points of refractory alloys, experimental measurements of melting properties are particularly challenging and not always available, as for the present TaVCrW alloy. Therefore, the melting temperatures for high entropy refractory alloys are often approximated as the mean value of the melting points of the constituting elements, which usually overestimates the actual experimental values~\cite{ZrNbTiVHf}. The proposed approach provides a computational performant alternative to such more coarse approximations. 

\section{Methods\label{steps}}
\subsection{Details of the methodology\label{method_detail}}

We sketch the entire workflow of the proposed method in five steps and compare them to those of the conventional TOR-TILD. 

\textbf{Step 1:} This step remains the same as in the conventional TOR-TILD. We apply the interface method~\cite{MorrisInterfaceMethod}, also called coexistence approach, to compute the melting temperature of the EAM, $T^{\rm m}_{\rm EAM}$. The employed interface structure is shown in Fig.~\ref{fig:master} (a). The efficiency of EAM allows us to perform calculations on sufficiently large supercells and long timescales. Here, a 16$\times$16$\times$32 supercell (16384 atoms) is used for TaVCrW with the simulation time up to 50 ps. The simulations are carried out with our previously developed automated pyiron workflow~\cite{zhu3}. As mentioned, determining the melting point of an empirical potential with high precision requires performing a few tens of calculations with different initial random seeds. The resulting statistical distribution of melting points closely resembles a Gaussian function, and the mean is used as a precise prediction, as shown in Fig.~\ref{fig:master} (a). The corresponding standard error is well below 1\,K and can be neglected in the final DFT melting point prediction.

At $T^{\rm m}_{\rm EAM}$ the volume of solid, $V^{\rm m,solid}_{\rm EAM}$, and the volume of liquid, $V^{\rm m,liquid}_{\rm EAM}$, are respectively computed by $NPT$ MD simulations using the EAM potential. The solid free energy $F^{\rm solid}_{\rm EAM}$($V^{\rm m,solid}_{\rm EAM}$,$T^{\rm m}_{\rm EAM}$) is then calculated by thermodynamic integration from an effective QH (computed in Ref.~[\onlinecite{Duff2}]) through
\begin{eqnarray}
F_{\rm EAM}^{\rm solid}(V^{\rm m,solid}_{\rm EAM},T^{\rm m}_{\rm EAM}) \:&=&\: F_{\rm QH}^{\rm solid}(V^{\rm m,solid}_{\rm EAM},T^{\rm m}_{\rm EAM})\nonumber\\
  &+& \vspace{-.3cm}\int_0^1\! d\lambda \left\langle E^{\rm solid}_{\rm EAM}-E^{\rm solid}_{\rm QH}\right\rangle_{\lambda,T^{\rm m}_{\rm EAM}},\nonumber\\
\end{eqnarray}
as shown in Fig.~\ref{fig:master} (b). Here, $\langle.\rangle_{\lambda,T}$ denotes the thermodynamic average at a temperature $T$ and coupling $\lambda$. A simple Einstein model is also tested as a reference. The free energy difference between using an effective QH and an Einstein model as reference is only 0.7 meV/atom, indicating that utilizing different references at this stage has a small impact on the computational accuracy. Once $F^{\rm solid}_{\rm EAM}$($V^{\rm m,solid}_{\rm EAM}$,$T^{\rm m}_{\rm EAM}$) is obtained, $F^{\rm liquid}_{\rm EAM}$($V^{\rm m,liquid}_{\rm EAM}$,$T^{\rm m}_{\rm EAM}$) is simultaneously achieved based on the relation $G(P,V)$\,=\,$F(V,T)$\,+\,$PV$, as the liquid Gibbs energy equals to the solid one at the melting point and constant pressure (here $P$\,=\,0\,GPa). Note that the melting properties at different pressures can be easily accessed by adding $PV$ term in the Helmholtz free energies.

\textbf{Step 2:} 
In this step using $F^{\rm liquid}_{\rm EAM}$($V^{\rm m,liquid}_{\rm EAM}$,$T^{\rm m}_{\rm EAM}$) from Step 1 as a starting point, the liquid free energy of MTP is obtained by performing thermodynamic integration through
\begin{eqnarray}
F_{\rm MTP}^{\rm liquid}(V^{\rm m,liquid}_{\rm EAM},T^{\rm m}_{\rm EAM}) \:&=&\: F_{\rm EAM}^{\rm liquid}(V^{\rm m,liquid}_{\rm EAM},T^{\rm m}_{\rm EAM})\nonumber\\
  &+& \vspace{-.3cm}\int_0^1\! d\lambda \left\langle E^{\rm liquid}_{\rm MTP}-E^{\rm liquid}_{\rm EAM}\right\rangle_{\lambda,T^{\rm m}_{\rm EAM}},\nonumber\\
\end{eqnarray}
as demonstrated in Fig.~\ref{fig:master} (c).  MTP is involved in this step compared to the conventional TOR-TILD using two EAMs.\par

\textbf{Step 3:} In this step, using $F^{\rm liquid}_{\rm MTP}$ ($V^{\rm m,liquid}_{\rm EAM}$, $T^{\rm m}_{\rm EAM}$) from Step 2 as a starting point, we map the liquid free energy surface of MTP by integrating the pressure $P(V,T)$ along the volume dimension using
\begin{equation}
\begin{aligned}
F_{\rm MTP}^{\rm liquid}(V,T^{\rm m}_{\rm EAM})\:=&\;F_{\rm MTP}^{\rm liquid}(V^{\rm m}_{\rm EAM},T^{\rm m}_{\rm EAM})\\[.06cm]&+\int_{V^{\rm m}_{\rm EAM}}^{V}\!\!\! P(V',T^{\rm m}_{\rm EAM})\,dV',
\label{eq:PV}
\end{aligned}
\end{equation}
and integrating the internal energy $U(V,T)$ along the temperatures dimension using 
\begin{equation}
\begin{aligned}
\frac{F_{\rm MTP}^{\rm liquid}(V,T)}{k_{\rm B}T}\:=&\:\frac{F_{\rm MTP}^{\rm liquid}(V,T^{\rm m}_{\rm EAM})}{k_{\rm B}T^{\rm m}_{\rm EAM}}\\[.06cm]&+\int_{T^{\rm m}_{\rm EAM}}^T \!\!d \left(\frac{1}{T'}\right)U(V,T'),
\label{eq:UT}
\end{aligned}
\end{equation}
where $k_{\rm B}$ is the Boltzmann constant. This step is illustrated in Fig.~\ref{fig:master} (d). Here, the MTP fully substitutes EAM compared to the conventional TOR-TILD. As applying MTP is much slower than using EAM, a relatively smaller but still converged supercell size of 12$\times$12$\times$12 (3456 atoms) is employed instead of a supercell size of 16$\times$16$\times$16 (8192 atoms) when using EAM.

\textbf{Step 4:}
In the conventional TOR-TILD method, this step is carried out by \textit{ab initio} MD simulations for thermodynamic integration calculations at various $\lambda$ values, volumes, and temperatures. Here, only classic MD simulations are carried out using the MTP to generate the trajectories at a set of volumes and temperatures (the black line in Fig.~\ref{fig:master} (e)). The applied supercell size is 4$\times$4$\times$4 (128 atoms) and the same as used for the DFT calculations in Step 5. Here, the computational effort is tremendously reduced by a magnitude of 10$^4$ in CPU core hours compared to the conventional TOR-TILD using EAM+MTP+TI, see Table~\ref{tab:CPU}.

\textbf{Step 5:}
This step aims to achieve the \textit{ab initio} accuracy. In the conventional TOR-TILD approach, free energy perturbation theory is indirectly applied, i.e., the up-sampling technique~\cite{UP-TILD}, by computing the energy difference between two sets of \textit{ab initio} energies separately calculated with low and high-converged DFT parameters. Here, we introduce a direct application of FEP, i.e., conducting static DFT calculations using high-converged DFT parameters on a set of uncorrelated snapshots from the MTP MD trajectories generated in Step 4. The resulting DFT energies are represented by the red dots in Fig.~\ref{fig:master} (e). The energy difference between the MTP and DFT can be calculated using Eq.(\ref{eq:FEP}). Importantly, this difference is very small, as evidenced by the narrow gap between the black and red dashed lines in Fig.\ref{fig:master} (e). This highlights the high quality of the MTP for performing free energy perturbation calculations.

It is crucial to note that the MTP is fitted to the DFT liquid energies without considering electronic contributions. Static DFT calculations must be performed at consistent electronic temperatures to fully incorporate electronic contributions, including vibration-electron coupling. The electronic free energy can be evaluated by a similar equation as Eq.~(\ref{eq:FEP})
\begin{equation}
\begin{aligned}
F^{\rm el}\!=-{k_{\rm B}T}{\rm ln}\langle e^{-\frac{1}{k_{\rm B}T}(E_{\rm el}^{\rm DFT}-E^{\rm DFT})}\rangle_{\rm MTP},
\end{aligned}
\label{eq:FEPFel}
\end{equation}
where $E^{\rm DFT}_{\rm el}$ and $E^{\rm DFT}$ are, respectively, the DFT energies including and excluding electronic temperature. At this end, the liquid free energy surface, including full vibration and electronic contributions, is obtained, as shown in Fig.~\ref{fig:master} (f).

\subsection{Potential fitting: EAM and MTP\label{PotentialFitting}}
In the proposed EAM+MTP+FEP approach, two potentials are fitted to different datasets for different purposes. We utilize the MEAMfit code~\cite{MEAMfit} for fitting the EAM potential. This code takes MD trajectories along with corresponding energies and/or forces as input. As the EAM potential serves solely to access the free energy of the MTP efficiently, it is fitted only on the solid DFT energies from \textit{ab initio} MD trajectories computed with low-converged DFT parameters (a cutoff energy of 300 eV and $k$-mesh with 2$\times$2$\times$2). Four volumes from the relevant volume range are selected for generating the training data at 2500\,K. This potential is used in Step 1 and Step 2. 

For fitting MTP, we employ the MLIP package~\cite{MLIP1} implemented in pyiron~\cite{pyiron} within the potential fitting module~\cite{container}. As an accurate MTP that can reproduce \textit{ab initio} liquid phase space is crucial for the efficiency of our proposed methodology, a two-stage fitting procedure is adopted. First, we fit a low-quality MTP with level 20 (lowMTP) to \textit{ab initio} liquid MD trajectories computed with low-converged DFT parameters, akin to those used for fitting the EAM. The fitting temperature is 2500\,K and five volumes from the relevant volume range are selected. The lowMTP is then used to generate new MD trajectories at the same temperature and volumes, followed by static DFT calculations with high-converged DFT parameters (a cutoff energy of 450 eV and $k$-mesh with 4$\times$4$\times$4) on a set of uncorrelated MTP MD snapshots. Second, we fit a high-quality MTP with level 24 (highMTP) using the DFT energies from the static DFT calculations in the preceding step. The RMSE of highMTP for GGA-PBE calculations is 2.6 meV/atom. Addressing the challenges outlined in Sec.~\ref{intro} concerning fitting a good reference for liquid multi-component materials, our highMTP for liquid TaCrVW demonstrates a sufficiently small RMSE in energies when compared to the MTP fitted for solid TaCrVW~\cite{Duff2} with an RMSE of 2.4 meV/atom. highMTP is used in Steps 2, 3 and 4.

\subsection{Short-range order calculations}
To model short-range order, we utilize the low-rank interatomic potentials (LRP){~\cite{shapeev2017,meshkov2019}} as an interaction model in canonical Monte Carlo (MC) simulations. The LRPs belong to a class of on-lattice machine-learning potentials and have been proven to perform very efficiently for computing SRO parameters in multi-component alloys{~\cite{PhysRevMaterials.5.053803, ghosh2022short, kostiuchenko2020short, kostiuchenko2019impact, ferrari2023simulating}}. We have evaluated different ranks and have chosen a rank of 3 for the final evaluation. The training has been performed on 128-atom cells by training a set of ten independent potentials to evaluate the uncertainty. During the iterative retraining process, 324 and 36 configurations have been added to the final training and validation sets, respectively. The training and validation errors are below 1 meV/atom. The Monte Carlo simulations have been carried out with periodic boundary conditions. We mainly focus on the 2000 -- 2400 K temperature range where the alloy remains disordered. The simulations are carried out for systems containing 128 and 5488 atoms, i.e., $4 \times 4 \times 4$ and $14 \times 14 \times 14$ lattice units, based on a two-atom primitive BCC cell. The number of MC steps (atom-atom swap attempts) is 20,000 times the number of atoms in the corresponding supercell. The burn-in approach{~\cite{cowles1996}} is utilized for thermal equilibration, i.e., for each temperature, the first half of MC steps was neglected. Test calculations with 50,000 MC steps per atom were performed to corroborate the findings. 
From the MC simulations, the internal energy and specific heat capacity are directly accessible. To compute the configurational contribution to the entropy, we follow the approach in Ref.~[\onlinecite{PhysRevMaterials.5.053803}], i.e., we integrated the specific heat capacity and subtracted it from a high-temperature reference state taken at 10,000\,K representing the ideal disordered state with entropy $\ln 4 k_B$. \par

The DFT calculations for the training have been performed at a lattice constant of 3.171\,{\AA} and using an electronic smearing parameter of 0.195 eV, which corresponds to the (GGA-PBE) values at 2260\,K, about 100\,K below the predicted melting temperature. The details of the other DFT parameters are as in Sec.~\ref{comp_details}. For computational efficiency, the plane wave cutoff and $k$ point mesh were reduced to 270 eV and  3$\times$3$\times$3, respectively. Test calculations from the training set revealed that more accurate parameters resulted in a constant energy shift with energy variation among the configurations below 0.2 meV/atom. 

\subsection{Computational details of the DFT and MD calculations\label{comp_details}}

\begin{table}
\begin{ruledtabular}
\caption{\label{tab:FittingSurfaceSOL}For solid TaVCrW from LDA calculations, the mesh of explicitly computed volumes, $V$ (per atom), and temperatures, $T$, that are used for obtaining the free energy surface. The volumes are additionally expressed in terms of a corresponding bcc lattice constant, $a=(2V)^{1/3}$.}
\begin{tabular}{cccc}
&$V$(\AA{}$^3$)  & 13.9, 14.19, 14.47, 14.75, 15.04\\
LDA&$a$(\AA{})  & 3.03, 3.05, 3.07, 3.09, 3.11 \\
&$T$(K)          & 2300, 2400, 2500, 2600, 2700, 2800 \\
\end{tabular}
\end{ruledtabular}
\end{table}

\begin{table}
\begin{ruledtabular}
\caption{\label{tab:FittingSurfaceLIQ}For liquid TaVCrW from GGA-PBE and LDA calculations, the mesh of explicitly computed volumes, $V$ (per atom), and temperatures, $T$, that are used for obtaining the DFT free energy surface. The volumes are additionally expressed in terms of a corresponding bcc lattice constant, $a=(2V)^{1/3}$.}
\begin{tabular}{cccc}
&$V$(\AA{}$^3$)  & 15.04, 15.33, 15.63, 15.93, 16.23 \\
GGA-PBE&$a$(\AA{})  & 3.11, 3.13, 3.15, 3.17, 3.19 \\
&$T$(K)          & 2200, 2400, 2600, 2800 \\
&$V$(\AA{}$^3$)  & 14.47, 14.75, 15.04, 15.33, 15.63\\
LDA&$a$(\AA{})  & 3.07, 3.09, 3.11, 3.13, 3.15 \\
&$T$(K)          & 2300, 2400, 2500, 2600, 2700, 2800 \\
\end{tabular}
\end{ruledtabular}
\end{table}

For the DFT calculations, we use the projector-augmented wave (PAW) method{~\cite{PAW}} as implemented in the VASP software package{~\cite{GK1,GK2,GK3,GK4}}. PAW potentials for Ta, Cr, V, and W treating $p$ electrons as the valence states are used. LDA and GGA are employed for the exchange-correlation functional, with the Perdew-Burke-Ernzerhof (PBE){~\cite{PBE}} parametrization for GGA. Note that the data of solid free energy from GGA-PBE are from Ref.~[\onlinecite{Duff2}]. In the present work, the liquid free energy of TaVCrW employing the proposed approach using both GGA-PBE and LDA, and the solid free energy employing the aforementioned direct upsampling approach using LDA are specifically calculated. The sets of explicitly DFT computed volume and temperature points for both solid and liquid in the present work are given in Table~\ref{tab:FittingSurfaceSOL} and Table~\ref{tab:FittingSurfaceLIQ}. The DFT calculations are performed in a 4$\times$4$\times$4 supercell with 128 atoms. For solid a 128-atom SQS structure is used. For liquid the SQS structure is heated up to a high temperature, e.g., 3000 K, to trigger the liquid phase. The explicitly computed DFT points are used as input to fit polynomials up to third order to obtain an analytical description of the free energy surface as a function of volume and temperature. The plane wave cutoff and $k$ point mesh (Monkhorst-Pack{~\cite{MPkmesh}}) are set to 450 eV and 4$\times$4$\times$4, respectively. For details on the fitting procedure of the corresponding GGA-PBE solid free energy surface and the respective convergence parameters, we refer to Ref.~[\onlinecite{Duff2}]. \par

For the reference potential calculations, we use the LAMMPS software package{~\cite{LAMMPS}}. For fitting the liquid free energy surface of our potential, we use the same volumes as for the DFT calculations (see Table~\ref{tab:FittingSurfaceLIQ}), but at a denser temperature mesh (steps of 20~K). For the MD simulations using the specially designed MTP we use a time step of 5 fs and the Langevin thermostat with a friction parameter of 0.01 fs$^{-1}$ to control the temperature.

\section{Acknowledgements}
L.-F.Z. acknowledges the funding by Deutsche Forschungsgemeinschaft (DFG, 493417040). F.K. and B.G. acknowledge funding from the European Research Council (ERC) under the European Union’s Horizon 2020 research and innovation programme (grant agreement No.~865855). F.K. acknowledges the LRP and MC simulation packages by Alexander Shapeev.

\section*{Data Availability}
The authors declare that the data supporting the findings of this study are available within the article and its supplementary information files or from the corresponding authors on reasonable request.

\section*{Author contributions}
L.F.Z. and B.G. designed the project. L.F.Z. acquired funding for this project, developed the method and performed the DFT calculations. F.K. performed SRO calculations. Q.C. and M.S. performed CALPHAD assessment. J.N. and B.G. provided valuable scientific insight and discussions. All authors discussed the results and contributed to writing the manuscript.

\section*{Competing interests}
The authors declare no competing interests.

\end{document}